\documentclass{article}
\pdfoutput=1
\usepackage{amsmath,amssymb,graphicx}
\usepackage{framed}
\def\be{\begin{eqnarray}}
\def\ee{\end{eqnarray}}

\def\tr{{\rm tr}\,}
\def\Tr{{\rm Tr}\,}

\def\l[{\phantom.[}

\hoffset= - 1.0in         

\begin{document}

\title{\vspace{.1cm}{\LARGE {\bf Towards topological quantum computer
}\vspace{.5cm}}
\author{
{\bf D. Melnikov$^{a,b}$},
\ {\bf A. Mironov$^{a,c,d}$},
\ {\bf S. Mironov$^{a,e}$},
\ {\bf A. Morozov$^{a}$},
\ {\bf An. Morozov$^{a,d,f}$}
}
\date{ }
}

\maketitle

\vspace{-6.2cm}

\begin{center}
\hfill FIAN/TD-02/17\\
\hfill IITP/TH-03/17\\
\hfill ITEP/TH-07/17\\
\hfill INR-TH-2017-006
\end{center}

\vspace{3.cm}

\begin{center}
$^a$ {\small {\it ITEP, Moscow 117218, Russia}}\\
$^b$ {\small {\it International
Institute of Physics - UFRN, Campus Universitario, Lagoa Nova, CP:
1613, Natal, RN 59078-970, Brazil}}\\
$^c$ {\small {\it Lebedev Physics Institute, Moscow 119991, Russia}}\\
$^d$ {\small {\it Institute for Information Transmission Problems, Moscow 127994, Russia}}\\
$^e$ {\small {\it Institute of Nuclear Research, Moscow 117312, Russia
  }}\\
$^f$ {\small {\it Laboratory of Quantum Topology, Chelyabinsk State University, Chelyabinsk 454001, Russia }}
\end{center}

\vspace{.5cm}

\begin{abstract}
One of the principal obstacles on the way to quantum computers
is the lack of distinguished basis in the space of unitary evolutions
and thus the lack of the commonly accepted set of basic operations
(universal gates).
A natural choice, however, is at hand: it is provided by
the quantum $R$-matrices, the entangling
deformations of non-entangling (classical) permutations,
distinguished from the points of view of group theory,
integrable systems and modern theory of non-perturbative calculations
in quantum field and string theory.
Observables in this case are (square modules of) the knot polynomials,
and their pronounced integrality properties could provide a key to
error correction.
We suggest to use $R$-matrices acting in the space of irreducible representations,
which are unitary for the real-valued couplings in Chern-Simons theory,
to build a topological version of quantum computing.
\end{abstract}

\bigskip

\bigskip

\paragraph{1. Introduction.}

Quantum computer\footnote{The literature devoted to quantum computing and related issues is incredibly vast, we just give text-book references \cite{Nielsen,QC} and a review \cite{Nayak}, further references can be found in these sources.} is proposed as a tool for solving some of the $NP$ problems outside of the class $P$ (if any), i.e. those
requiring exponentially large resources, if conventional (classical)
computers are used.
A typical example of such a problem is multiplication of exponentially large
matrices, and the key observation is that Nature provides a solution to it
at no expense.
Indeed, a system of $M$ spins has a Hilbert space of dimension $2^M$
and evolution operator is a unitary matrix of size $2^M\times 2^M$:
thus, these huge evolution matrices are automatically multiplied (iterated)
as time goes.
The task of {\it quantum computing} is a reduction of problems to
multiplication of huge unitary matrices, which becomes a straightforwardly
solvable problem once quantum computer gets available.

Quantum programming is a deformation and extension of classical reversible programming. This later represents any algorithm as a series of permutations which can be realized by a corresponding product of permutation matrices.
In order to make a reversible algorithm, one avoids erasing, copying and forgetting operations,
which change the entropy
and cause unavoidable heating, at least, by $bit$ costs $kT\log 2$ (von Neumann-Landauer limit) \cite{NL}.
Any classical algorithm can be reformulated reversibly,
though it is not always a trivial task (in particular, it requires, at least, 3-bit universal gates).
Quantum programming substitutes
permutations by generic unitary transformations,
thus enormously extending the set of reversible algorithms,
what, among other things, may solve some $NP$-class problems
at the expense of providing the answers only probabilistically
(what is not a practical limitation, since testing the answer requires a polynomial time for the $NP$-class problem).
The restriction to the {\it unitary} matrices is partly artificial,
the reason is just that we know the physical mechanism
(quantum mechanics) for multiplication of unitary, but not any other
types of matrices.
If there will be a realization of multiplication of any other
matrices, it would provide
a reasonable alternative for the current quantum computer paradigm.

In fact, quantum computing is not yet universally defined,
it is even not yet decided whether accent should be put
on "analogous" (system oriented) or on "digital" devices.
Different thinkable processors use different sets of universal gates,
and their choice can seriously affect the actual programming.

A new twist in this story is provided by the idea \cite{Kauf} of
{\it topological} programming,
when permutations are substituted by quantum $R$-matrices, which are the natural deformations of permutations, on one hand, and are related to topological (and integrable) structures, on the other hand. These topological structures are encoded in the corresponding knot/link polynomials.
Needed are permutations of arbitrary size and they are substituted
by $R$-matrices in higher representations which in the case of links
can act on the products of not necessarily coinciding representations.

To begin with, there are serious chances that knot theory will finally provide
a method of calculating knot polynomials not directly, by multiplication of
huge $R$-matrices, but via discovery
of their properties like the modified Reshetikhin-Turaev formalism \cite{RTmod,inds},
additional equations \cite{Gar,IMMMfe}, simple recursions \cite{MMpol}, hidden dualities and integrabilities \cite{MMMI}.
If this happens, one would have a solution to an $NP$ problem, of which
the knot polynomial calculus, if approached directly, is a typical example.

More conservative viewpoint is, however, that the knot polynomials
are a typical problem to be solved with the help of quantum computers.
And it is clearly a distinguished problem, what makes the topological computer,
targeted to (best adjusted for) solving this particular problem a
distinguished candidate for the role of the standard setting machine
in the field, an analogue of the Turing machine in classical programming.
Perhaps not accidentally, there is a built-in bonus in the topological approach:
in the case of knots the observables are {\it polynomials}
(in the coupling constants $q$ and $A$),
moreover they have {\it integer} coefficients,
and this provides an easy way to test the validity of calculations
and thus a new fundamental tool for the error control in quantum computations.

There are two basic problems.
The theoretical one: the $R$-matrices are not necessarily unitary.
The "practical" one: it is difficult to realize $R$-matrices in a physical system,
normally this requires moving of twist operators in WZNW model \cite{AG}, which is much similar to the quantum computer realization (see the review in \cite{Nayak}), though one can also hope for a realization in terms
of Chern-Simons theory dynamics \cite{Freedman,Kitaev}.

Because of the recent progress in advanced
knot theory, theoretical problems can be now application-independent addressed in this theory, while
practical problems are best understood in modern considerations of
the Fractional Quantum Hall Effect.
Practical solutions should depend on the resolution of "theoretical" issues:
what kind of evolutions we actually wish to realize.
Here is one of the bifurcation points:
there are artificial choices of "accidentally" unitary $R$-matrices
and there is a {\it natural} class of those: in particular channels
obtained by projecting on irreducible representations (irreps).
In this note, we briefly discuss basics of the "theoretical" part.

{\bf The main message is that there is a distinguished choice of
quantum programming, the topological one based on $R$-matrices in irreducible representations,
and one should look for {\it their} realizations in systems,
which admit a non-Abelian WZNW description.}
Such quantum computers will be "analogous" for the most fundamental
archetypical quantum objects, colored knot polynomials \cite{knotpols,Wit,inds,RT,RTmod},
and this is the language which is best suited for formulating the
quantum programming in a universal, application independent way.
There are a lot of quantum evolutions, while what we need is the "fundamental" one, and it is provided by the knot polynomial calculus.

In a more formal language, one can say that the standard quantum programming is physically associated with the quantum entanglement, i.e. with EPR like effects, while the topological quantum programming deals with the topological entanglement, i.e. with topological linking. The idea to relate these two a priori different entanglements goes back to P.K.~Aravind \cite{Ara} and was later developed by L.~Kauffman and other authors, who suggested to use elements of the Hecke \cite{Hecke} (or the Temperley-Lieb \cite{TL}) algebras as realizations of quantum unitary operations \cite{Kauff2}. However, this algebra is too small, one needs to have much more to realize a full-fledged quantum programming. In this paper, we propose the set of $R$-matrices realized as acting in the space of intertwining operators in various representations as a possible realization of the universal gate sets for the purposes of quantum programming, i.e. to remove an artificial restriction to the fundamental representations of $SU(2)$, which unreasonably reduces braid groups to the Hecke algebras. Within this framework, the quantum codes are nothing but the quantum (graded) traces or matrix elements of products of these R-matrices, i.e. their role is actually played by the knot/link polynomials. Thus, in this way one can naturally associate  the sequence of quantum operations (gates) with a knot/link.

\paragraph{2. Quantum programming.}

Obviously, topological programming (TP) is intimately related to knot theory,
especially in the Reshetikhin-Turaev (RT) formalism \cite{RT},
and addressing the TP issues one can make use of the recent advances
in the theory of knot polynomials.

One of the first results in this direction is a recently proved connection between quantum and topological entanglements \cite{ten}.
The main problem with TP is that the group theory $R$-matrices
are not unitary at $|q|=1$. For instance, the simplest $R$-matrix (acting in the product of two $SU(2)$ fundamental representations) \cite{YB}
\be\label{Rf}
R={1\over q}\left(
\begin{array}{cccc}
q&&&\\
&\{q\}&1&\\
&1&0&\\
&&&q
\end{array}
\right)
\ee
where $\{x\}\equiv x-x^{-1}$, is already non-unitary at unimodular $q$.
There are two obvious ways out of this problem.

One is to make use of the unitary solutions to Yang-Baxter equations,
e.g. provided by representations of the Hecke \cite{Hecke} (Temperley-Lieb \cite{TL}) algebras
which govern the physics of spin chains (like the XXZ model \cite{sch,KBI}) and loop models \cite{TL,loop}.
This is a very interesting approach, in knot theory it is related to
study of the Khovanov formalism \cite{KhR} and its various modifications \cite{DM}.

Another possibility is to note that the group theory ${\cal R}$-matrices become
unitary when acting on irreducible representations, i.e. in the
modern version of the RT formalism \cite{inds,RTmod}.
One can fully use this way the above mentioned achievements of
modern knot theory.
Here the sacrifice is relation to the physical intuition about the spin
chains and, more than that, to the corresponding intuition about the quantum
evolution.
The point is that the states in different representations of quantum groups
are not orthogonal in the ordinary quantum mechanics, e.g.
\be
q|\uparrow\downarrow> + |\downarrow\uparrow>
\ \ \ \ {\rm and} \ \ \ \  |\uparrow\downarrow> - q|\downarrow\uparrow>
\ee
look orthogonal at real $q$, but not such at the unimodular $q$,
which is needed for unitarity.
This is because the scalar product involves complex conjugation.
In other words, we deal with the unitary evolution of two non-orthogonal states,
which can not be unified into a common evolution:
it can not be unitarily lifted from irreducible representations to
the full Hilbert space of the spin chain.
Still, we suggest to build a version of the TP on this unusual grounds.

\paragraph{3. $R$-matrices.}

Let us note that there are two different $R$-matrices that differ by the permutation of two spaces on which the $R$-matrix acts \cite{J,KBI}. One of the $R$-matrices is obtained from the universal $R$-matrix in the concrete representations and satisfies the Yang-Baxter equation \cite{YB} (in this letter we consider only constant $R$-matrices, since they are associated with knots, \cite[s.12]{DIM})
\be
\check R_{12}\check R_{13}\check R_{23}=\check R_{23}\check R_{13}\check R_{12}
\ee
where the indices denote the spaces where the $R$-matrix acts. For the fundamental representations and the group $SU(2)$, this $R$-matrix is
\be
\check R=\left(
\begin{array}{cccc}
q&&&\\
&1&\{q\}&\\
&0&1&\\
&&&q
\end{array}
\right)
\ee
The second $R$-matrix (\ref{Rf}) is obtained from this one by the permutation of columns and is related to constructing knot invariants via the Reshetikhin-Turaev construction \cite{RT}. This $R$-matrix satisfies the knot type Yang-Baxter equation
\be
R_{12}R_{23}R_{12}=R_{23}R_{12}R_{23}
\ee

The main drawback of the $R$-matrix (\ref{Rf}) for our needs is that it is not unitary for $|q|=1$, while for the quantum computing we need unitary operations. Instead, it is Hermitian for real $q$, this is why it appears as a Hamiltonian in the theory of spin chains. However, we want to use the $R$-matrices in the role of unitary evolution operators, not of Hermitian Hamiltonians. To make this possible, note that the $R$-matrix (\ref{Rf}) acts in the product of two fundamental representations of $SU(2)$. This product can be decomposed into the sum of two irreps: $[1]\otimes [1]=[2]+[11]$, symmetric and antisymmetric. An essential feature of the $R$-matrix is that it is constant on the whole irreducible representation. This is because the $R$-matrix commutes with the co-product (in variance with $\check R$). For instance, the eigenvalues of the $R$-matrix (\ref{Rf}) are $q,q,q,-1/q$. The first three eigenvalues correspond to the 3-dimensional spin 1 irrep $[2]$, while the fourth one is associated with the scalar (spin 0) irrep $[11]$. Hence, in the two-dimensional space of representations (or, better to say, intertwining operators) the $R$-matrix can be written as
\be\label{R1}
{\cal R}=\left(\begin{array}{cc}
q&0\\
0&-{1\over q}
\end{array}\right)
\ee
This matrix is unitary if $|q|=1$.

More generally, if one considers the $R$-matrix acting in the product of two representations $V_1$ and $V_2$ so that
\be\label{dec}
V_1\otimes V_2=\oplus_Q {\cal M}_{V_1V_2}^Q Q
\ee
where some irreps $Q$ can appear a few times at the r.h.s., the $R$-matrix can be realized in the space of all possible intertwining operators ${\cal M}$, the eigenvalue of $R$-matrix on the space $Q$ being equal (up to a sign) to $q^{\sum_{i,j\in Q}(i-j)}$, where the sum runs over the Young diagram describing the representation $Q$. The subtlety is, however, that, in the case of the irrep appearing several times at the r.h.s. of (\ref{dec}) the matrix ${\cal R}$ is no longer diagonal. Consider, for instance, the product of three fundamental representations. The $R$-matrix acting on the first two of them, ${\cal R}_1$ is (\ref{R1}), and such as the $R$-matrix ${\cal R}_2$ acting on the second and the third representations. However, these two $R$-matrices are diagonal in different bases. Let us rewrite them in the basis of irreps of the triple: $[1]^3=[3]+2[21]+[111]$. One can see that if the basis is chosen so that ${\cal R}_1$ is diagonal:
\be
{\cal R}_1=\left(\begin{array}{cccc}
q&&&\\
&q&&\\
&&-{1\over q}&\\
&&&-{1\over q}
\end{array}\right)
\ee
then
\be
{\cal R}_2=\left(\begin{array}{cccc}
q&&&\\
&-{1\over q^2[2]}&{\sqrt{[3]}\over [2]}&\\
&{\sqrt{[3]}\over [2]}&{q^2\over [2]}&\\
&&&-{1\over q}
\end{array}\right)
\ee
where $[n]\equiv \{q^n\}/\{q\}$ are the quantum numbers.
We will be interested only in the non-trivial $2\times 2$ part of the matrices related to the two irreps $[21]$, i.e.
\be
{\cal R}_1=\left(\begin{array}{cc}
q&0\\
0&-{1\over q}
\end{array}\right),\ \ \ \ \ \ \
{\cal R}_2=\left(\begin{array}{cc}
-{1\over q^2[2]}&{\sqrt{[3]}\over [2]}\\
{\sqrt{[3]}\over [2]}&{q^2\over [2]}
\end{array}\right)
\ee
and ${\cal R}_2$ here is obtained from ${\cal R}_1$ by the rotation of basis provided by the Racah matrix
\be\label{Racah2}
{\cal S}=\left(\begin{array}{cc}
{1\over [2]}&{\sqrt{[3]}\over [2]}\\
{\sqrt{[3]}\over [2]}&-{1\over [2]}
\end{array}\right)
\ee
Note that both ${\cal R}_1$ and ${\cal R}_2$ are unitary at unimodular $q$. The reason is that the matrix elements of Racah matrices are always constructed only from the quantum numbers, and these later are real at unimodular $q$. These formulas are immediately generalized to $SU(N)$ so that the $R$-matrices do not change.

Hence, one can generate large enough set of unitary operations, considering matrices ${\cal R}_i$ for large enough representations and acting on the product of sufficiently many spaces. We consider examples in the next paragraph.

\paragraph{4. Knot polynomials and closed braids.}

Having the $R$-matrix, one can construct the knot HOMFLY polynomial colored with a representation $V$ for an arbitrary knot by representing it as a closed braid so that, for the $n$ strand braid, the braid generator $b_i$ is associated with the $R$-matrix acting at $i$-th and $(i+1)$-th factors at the tensor product
\be
V_1\otimes\ldots\otimes V_{i-1}\otimes \underbrace{V_i\otimes V_{i+1}}_{R_i}\otimes V_{i+2}\otimes\ldots\otimes V_n
\ee
In other words, with each crossing at the two-dimensional projection of the knot one associates either the $R$-matrix or its inverse, depending on the order of strands at the crossing. One can similarly deal with links with different components colored with different representations. In this case, one sometimes deals with $R$-matrices acting on the product of different representations.

Thus, any knot/link is described by a product of $R$-matrices. The second essential ingredient in order to construct the knot/link polynomial is to define the trace of these $R$-matrices (since the braid is closed). In fact, one can calculate the standard quantum trace, i.e. the ordinary trace with inserted $q^{\rho(V^{\otimes n})}$, where $\rho$ is the Weyl vector (half-sum of all positive roots). This quantum trace reduces to the trace over each concrete irrep and the sum of the irreps, and the HOMFLY polynomial is
\be
H_V=\hbox{Tr}_{V^{\otimes n}} \left(q^{\rho(V^{\otimes n})}\prod R\right)=\sum_Q \hbox{tr}_Q \left(q^{\rho(Q)}\prod_{{along}\atop{braid}} R\right)
\ee
In the basis of intertwining operators discussed in the previous paragraph, the $R$-matrices are rewritten in terms of matrices ${\cal R}_i$ that are constant on irreps $Q$, i.e.
\be
H_V=\sum_Q \left(\prod {\cal R}\right)\left(\hbox{tr}_Q\, q^{\rho(Q)}\right)=\sum_Q d_Q \prod_{{along}\atop{braid}} {\cal R}
\ee
since $\hbox{tr}_Q\, q^{\rho(Q)}$ is equal to the quantum dimension $d_Q$ of the representation $Q$.

In more physical terms, one can associate these HOMFLY polynomials with the Wilson loop averages in Chern-Simons gauge theory \cite{CS} where the Wilson loop is just the knot/link \cite{Wit}. If one considers the gauge group $SU(N)$, the representation of the Wilson loop $V$ and the Chern-Simons coupling $\kappa$, the Wilson loop average is exactly the HOMFLY polynomial $H_V$, which is a Laurent polynomial (up to a normalization factor) of $q=e^{2\pi i\over \kappa +N}$ and $A=q^N$. This implies that, in the unitary theory, when $\kappa$ is real, $q$ is unimodular. Another interesting restriction is to consider an integer $\kappa$. This case describes integrable representations of the $SU(N)$ WZNW theory \cite{CFTKM}, the evolution of its conformal blocks being associated with the corresponding Wilson averages in the $SU(N)$ Chern-Simons theory \cite{Wit,CSmore}. In more practical terms, the integer $\kappa$ greatly simplifies calculations \cite{AS}, while, on the quantum computing side, it gives additional tools for {\it correcting errors}.

\paragraph{5. Knot polynomials and plat (multi-bridge) representations.}

There is another way to obtain the HOMFLY polynomials from the sequence of $R$-matrices, which is closer to the original suggestion by E.Witten \cite{Wit} and to physical realization of quantum calculations. That is, instead of taking trace, one can calculate vacuum matrix elements (or taking projections onto singlet representations) of the product of the same $R$-matrices in the space of intertwining operators \cite{inds,GMM}.

Within this approach, one has to consider the plat (or bridge) representation of the (oriented) knot/link, i.e. a braid with an even number of (oriented) strands so that the ends of strands are pairwise contracted (see Fig.1). Fixing positive direction along the braid, one can associate with the parallel strand the representation $V$ and with the antiparallel strand the conjugated representation $\bar V$. Since in the product $V\otimes\bar V$ there is the singlet representation, one can always project the unitary operator describing this braid to pairwise singlet representations (which is the exact meaning of contracting the ends).

\begin{figure}[h]
  \centering
\includegraphics[width=5cm]{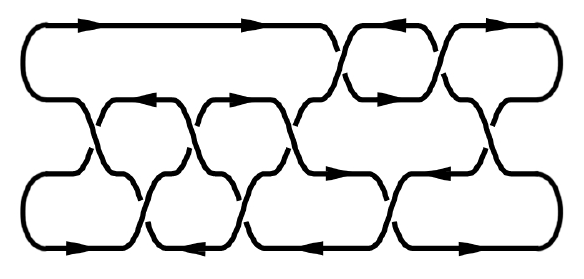}
\caption{Typical plat representation for the knot (it is the two-bridge case: there are two arcs at the left and at the right). This concrete knot is $9_{14}$ in the Rolfsen table \cite{katlas}.}
  \label{fig:3}
\end{figure}

Since in Fig.1 there are both parallel and antiparallel strands at any triple of strands, now one needs more ample set of the Racah matrices. Say, even in the simplest 4-stand case of the two-bridge knots depicted in the figure, there are two $R$-matrices
\begin{eqnarray}
{\cal R} = -\frac{1}{qA}\left(\begin{array}{cc} -q^2&0\\ 0 & 1 \end{array}\right)\ ,
\ \ \ \ \ \ \
\bar {\cal R} = \left(\begin{array}{cc} -A&0\\ 0 & 1 \end{array}\right)
\label{R}
\end{eqnarray}
and two different Racah matrices, depending on the mutual orientation of strands (we refer the reader to \cite{inds,GMM})
\begin{eqnarray}
{\cal S}= \frac{1}{[N]}\left(\begin{array}{cc} \sqrt{\frac{[N][N-1]}{[2]}} & \sqrt{\frac{[N][N+1]}{[2]}} \\ \\
\sqrt{\frac{[N][N+1]}{[2]}} & -\sqrt{\frac{[N][N-1]}{[2]}} \end{array}\right)\ ,\ \ \ \ \ \ \
\bar {\cal S} = \frac{1}{[N]}\left(\begin{array}{cc} 1 &  \sqrt{[N-1][N+1]}  \\ \\
\sqrt{[N-1][N+1]}  & -1 \end{array}\right)\label{S}
\end{eqnarray}
where $A=q^N$ and we changed the framing as compared with the previous paragraph. Note that in the case of $N=2$, i.e. $A=q^2$ the Racah and $R$-matrices pairwise coincide (up to the framing), since in the $SU(2)$ case the representation coincide with its conjugate.
These Racah matrices are unitary only with the additional constraint $[N\pm 1]\ge 0$, $[N]\ge 0$ (this is because of the square roots), which is the case for a sufficiently small phase of $q$.

If looking at Fig.1 as a four strand braid, one has to associate with the first strand ${\cal R}_1={\cal R}$ or ${\cal R}_1=\bar {\cal R}$, depending on the orientation of strands. Similarly, ${\cal R}_2={\cal S}{\cal R}{\cal S}^t$, ${\cal R}_2=\bar{\cal S}\bar{\cal R}{\cal S}$ or its transposed, depending on the orientation of strands. The proper choice is fixed by the matrix indices, which take values: $X\in V\otimes V$ and $\bar X\in\otimes V\otimes\bar V$. With the indices, the matrices are: ${\cal S}_{\bar X Y}$, $\bar{\cal S}_{\bar X\bar Y}$, ${\cal R}_{XY}$, $\bar{\cal R}_{\bar X\bar Y}$. At last, the third matrix ${\cal R}_3={\cal R}_1$ in this case, since the only representation in the product $V^{\otimes 4}$ that contributes to the necessary matrix element is singlet.

The HOMFLY polynomial corresponding to knot $9_{14}$ in Fig.1 is then given by the matrix element $(11)$ of the product
\be
H^{9_{14}}_V=d_V\cdot\left({\cal S}{\cal R}{\cal S}^t \bar {\cal R}^{-1}\bar {\cal S}\bar {\cal R}{\cal S}{\cal R}^{-1}{\cal S}^t \bar{\cal R}\bar {\cal S}\bar {\cal R}^{-3}{\cal S}{\cal R}{\cal S}^t \right)_{11}
\ee
where the superscript $t$ denotes transposition of the matrix. In the fundamental representation $V=[1]$, one can use the matrices (\ref{S}) and (\ref{R}) in order to produce the standard HOMFLY polynomial for knot $9_{14}$ (see \cite{katlas}).

Hence, the HOMFLY polynomials are matrix elements of unitary operators (up to a normalization factor) and
topological computing actually deals with evaluation of the absolute value of the HOMFLY polynomials, which
are less or equal to unity (in the range of values where the Racah matrices are unitary like $[N\pm 1]\ge 0$ in the fundamental case above).

Note that within this approach one starts from the $R$-matrices in the space of intertwining operators from the very beginning. This is because this construction is equivalent to dealing with monodromies of the conformal blocks of the WZNW model \cite{Wit} so that all the $R$-matrices act in the space of the conformal blocks. Thus, in this part, the approach is very close to that described in the review \cite{Nayak}, where a quantum computer is realized by moving points in the conformal blocks.

\paragraph{6. Versatility.}

As any programming, the one for quantum computers builds arbitrary algorithms
from a few standard operations called {\it universal gates}.
They realize a set of unitary $d^M\times d^M$ matrices, which, when considered as
multiplicative generators, form a subgroup, which is dense in the
entire set $U(d^M)$ of unitary matrices of a given size.
There is still no fully conventional choice of universal gates,
only some theoretical criteria for making this choice in the future.

In fact, an arbitrary product of matrices from the set of $k$ unitary ${\cal N}\times {\cal N}$ matrices $U_j=e^{iA_j}$,
$j=1,\ldots,k$,  can be given
by an infinite integer-valued  word $\{n_{aj}\}$,
\be
U_1^{n_{11}} \ldots U_k^{n_{1k}} U_1^{n_{21}}\ldots U_k^{n_{2k}} \ldots
= \prod_a \prod_{j=1}^k U_j^{n_{aj}}
\ee
and the Baker-Campbell-Hausdorff formula
expresses it as an exponential of a sum of repeated commutators of $A_j$
with rational $\{n\}$-dependent coefficients.
In order to approximate an arbitrary unitary matrix in this way, one needs to fulfil
three properties:

(a) there are  ${\cal N}^2$ linearly independent repeated commutators
of $k$ matrices $A_j$

(b) the $\{n\}$-dependent coefficients in front of them are all independent and

(c) the commutators fill densely the ${\cal N}^2$ plane, i.e. among repeated commutators
there are {\it infinitely many} linearly independent
modulo $Z$ (i.e. with {\it rational} coefficients) and the rank
is also unrestricted.

The property (a) means that there is nothing like Serre relations,
which restrict the growth of space of repeated commutators, in particular that
the matrices $A_j$ are {\it not} generated by the comultiplication from any small
Lie algebra, smaller than $SU({\cal N})$.
The properties (b) and (c) can be true, if the matrices $A_j$ are
themselves are irrational, then linear combinations with integer coefficients
can still be dense.
It is important that just the lack of Serre identities and irrationality
allows one to span the space of exponentially large, ${\cal N}\sim d^M$ ($M$ qudits) matrices
with just a few matrices. These matrices, as was explained above, can not be subject to Serre relations in some $SU(k)$ subgroup.

In fact,  just two matrices are sufficient to approximate {\it any} unitary.
For example, one can take for these two $\exp\Big(\sum_i c^+_iF_i\Big)$ and $\exp\Big(\sum_i c^-_iE_i\Big)$, where $c^{\pm}_i$ are arbitrary irrational coefficients such that their ratios are also irrational, and $F_i$ ($E_i$) are the Chevalley generators corresponding to positive (negative) simple roots of $SU({\cal N})$ in the fundamental representation.
We provide a simple illustration of how this works in the case of $SU({\cal N})=SU(4)$
First, take two elements $A_1$ and $A_2$ in the Lie algebra that do not lie in a smaller subalgebra, i.e. which can generate the whole algebra (the everywhere dense subset in it)
after repeated commutating.
We choose
\be
A_1={E_1\over\sqrt{3}}+{E_2\over\sqrt{5}}+{E_3\over\sqrt{7}},\ \ \ \ \ \ \ \ A_2={F_1\over\sqrt{11}}+{F_2\over\sqrt{13}}+{F_3\over\sqrt{17}}
\ee
where $F_i$ ($E_i$) correspond to positive (negative) Chevalley generators in the fundamental representation.
Second, we repeatedly commute $A_1$ and $A_2$, take the linear span with integer coefficients of the obtained commutators and check that it densely covers the whole positive sub-space of $SU(4)$ algebra (for the compact group this positivity at the algebra level is not a restriction).
In fact, one do not even need {\it all} the repeated commutators:
a rather particular sequence is actually enough.

Let us denote $\alpha=[[A_1,A_2],A_1]$ and $\beta=[[A_1-3\alpha,A_2],A_1]$ (technically convenient is to have $\alpha_{12}\gg\alpha_{23}$ and $\beta_{12}\ll\beta_{23}$). Now we start with $A_1$, commute it with $A_2$, then with $A_1$, then add $\beta$ with an integer coefficient, commute it again first with $A_2$ and then with $A_1$ etc. Thus, we define the sequence
\be
B[i] \rightarrow [[B[i],A_2],A_1] \rightarrow B[i]+p^{(1)}_i\alpha+p^{(2)}_i\beta=B[i+1]
\ee
The integers $p^{(1)}_i$ and $p^{(1)}_i$ are random, but chosen so that to keep
the coefficients in front of $E_1$ and $E_2$ within the segment $[0,1]$. Due to this choice, without adding $\alpha$ and $\beta$, all coefficients would tend to zero. The picture generated by the computer is depicted in Fig.2.

\begin{figure}[h]
  \centering
\includegraphics[width=5cm]{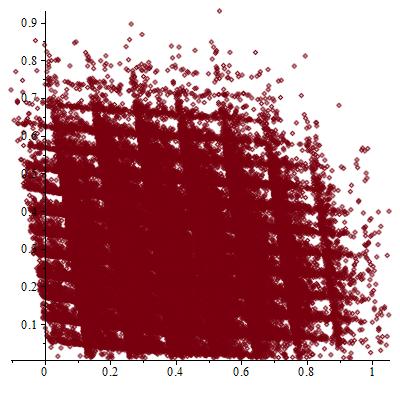}
\caption{The plot of coefficients in front of generators $E_1$ and $E_2$.}
  \label{fig:4}
\end{figure}

\noindent

The  periodic structure, which is clearly seen in the plot,
emerges due to a particular regular procedure that has been used,
still  the background is densely filled.

One can illustrate the same phenomenon immediately in the group terms: let us take two randomly chosen unitary $4\times 4$ matrices $U_1$ and $U_2$. Now one can check if they generate all the unitary matrices just via generation of the random sequences $\{n_i\}$ and then looking at the matrix elements of $U=U_1^{n_1}U_2^{n_2}U_1^{n_3}\ldots$. A typical plot obtained this way is depicted in Fig.3.

\begin{figure}[h]
  \centering
\includegraphics[width=5cm]{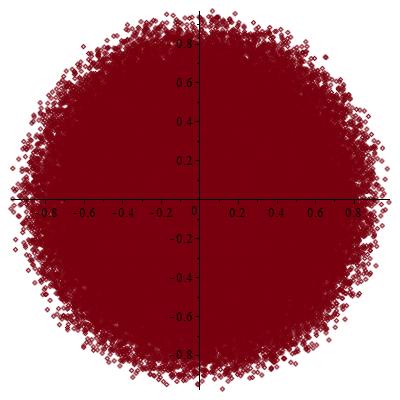}\hspace{3cm}
\includegraphics[width=5cm]{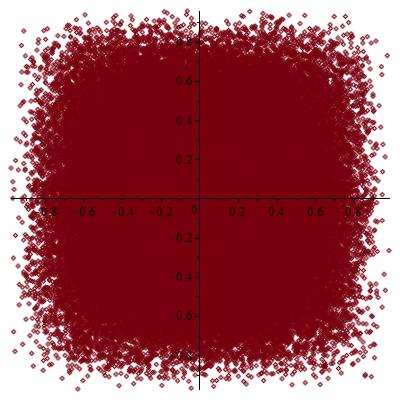}
\caption{The plot of the matrix elements of the unitary matrix randomly generated from two random unitary matrices: $\hbox{Re } U_{11}$ and $\hbox{Im } U_{11}$ (left); $\hbox{Re } U_{11}$ and $\hbox{Re } U_{23}$ (right).}
  \label{fig:4}
\end{figure}

\noindent

However, one can also see that the plots are dense, but not-homogeneous,
thus different unitary matrices can be approximated with different effort
(at different cost).
This once again emphasizes the advantage of choosing the generators not randomly,
but "cleverly", so that the really important codes are easier accessible.
The hope expressed in this paper is that ${\cal R}$-matrices can provide
such a "clever" choice for most fundamental problems.

\paragraph{7. Universal gates.}
In fact, the two generators, though enough to approximate an arbitrary unitary matrix,
do not have a "local" structure and, hence, are not the true universal gates needed in quantum programming, where one wants to realize the whole code, acting on exponentially large matrices,  in terms of a set of small standard qubits (or qudits).
In other words, it is not sufficient to an approximate arbitrary ${\cal N}\times{\cal N}$
unitary matrix, one should also respect the auxiliary structure reflected in the fact that
${\cal N}=d^M$. Moreover, the emerging
generators should be {\it local}, but {\it not}  built by comultiplication from the $d\times d$ ones, exact requirement still needs to be formulated in abstract group theory terms. In practice, for the qubit realization and ${\cal N} = 2^M$, the minimal necessary number of universal (i.e. "local" and "quasi-local") gates is believed to be $3M-1$.

For the simplest case of the system of two qubits, i.e. $d=2$, $M=2$
the typical set of universal gates
consists of five matrices. One of them is necessarily an {\it entangling} matrix \cite{Br},
the others are chosen in the form of single-qubit evolutions, $U\otimes I$ and $I\otimes U$
which are non-entangling. For instance, the {\it standard} universal set consists of
CNOT and single qubit Hadamard and $\pi/8$ gates (so that on two qubits the matrices are $I\otimes H$, $H\otimes I$, $I\otimes T$ and $T\otimes I$):
\be
\hbox{controlled-NOT}:\ \ \ K=\left(\begin{array}{cccc} 1 \\ & 1 \\ &&&1 \\ && 1 \end{array}\right),
\ee
\be\label{Hadamard}
\hbox{Hadamard}:\ \ \ H=\frac{1}{\sqrt{2}}\left(\begin{array}{cc} 1&1\\ 1 & -1 \end{array}\right),
\ee
\be\label{Pgate}
\pi/8\hbox{-gate}:\ \ \ T=\left(\begin{array}{cc} 1&0\\ 0 & e^{i\pi/4} \end{array}\right)
\ee
One can prove that any single qubit unitary operation can be approximated by the Hadamard and $\pi/8$ gates to arbitrary accuracy. Now, the universality of the above set is guaranteed by the Brylinskis theorem \cite{Br}, which claims that the two-qubit gate $G$ along with local unitary gates (in our example, Hadamard and $\pi/8$ gates) provides the universal set of gates if and only if $G$ is entangling, i.e. if there exists a vector $|v>=|x>\otimes |y>\in V_2\otimes V_2$ such that $G|v>$ is not decomposable as a tensor product of two qubits.

The proof consists of a few steps (for an alternative proof see \cite{Nielsen,Br}). First of all, one proves that any unitary $2\times 2$ matrix can be realized up to arbitrary accuracy by a product of matrices $H$ and $T$. This is done by noting that generating all such matrices can be reached by two different rotation generators on the Bloch sphere $R_{\vec n}(\theta)=\exp\Big(-i\theta\vec n\cdot\vec\sigma/2\Big)$ ($\sigma_i$ are the Pauli matrices) with non-parallel $\vec n_1$ and $\vec n_2$ and non-rational angles $\theta_{1,2}$. $HTHT$ and $THTH$ provide an example of such generators. The next step is to note that any $4\times 4$ unitary matrix can be presented in the form
\be\label{um4}
U_{4}=\exp \Big(\sum_{i_j=0}^3 c_{ij}\, \sigma_{i}\otimes\sigma_{j}\Big)
\ee
where $\sigma_0=I$. Now, since $Ue^AU^{-1}=e^{UAU^{-1}}$, one can apply the single qubit matrices $H$ and $T$ to the first and the second factor of the tensor product in order to transform to each other linear combinations of the Pauli matrices in (\ref{um4}). The only problem is to generate from $e^{\sigma\otimes I}$ some $e^{\sigma\otimes\sigma}$. This is done by the CNOT (or any other entangling) matrix: $Ke^{\sigma_1\otimes I}K=e^{\sigma_1\otimes\sigma_1}$. This completes the construction.

This scheme is immediately generalized to multi-qubit case, when, in order to approximate {\it arbitrary} $2^M\times 2^M$ unitary matrices, one needs just $2M$ single qubit matrices made of $H$ and $T$, and a few entangling matrices. Indeed, in this case,
\be
U_{2^M}=\exp \Big(\sum_{i_j=0}^3 c_{\vec i}\,\sigma_{i_1}\otimes
\ldots \otimes\sigma_{i_M}\Big)
\ee
and this time one needs $M-1$ entangling matrices to deal with tensor products containing $\sigma_0=I$. For instance, one can choose $N-1$ CNOT matrices $K_i$ acting on the $i$-th factors: $K_i (I\otimes \ldots \otimes  \sigma_1\otimes I\otimes\ldots\otimes I)K_i=I\otimes \ldots \otimes \sigma_1\otimes\sigma_1\otimes\ldots\otimes I$. Thus, totally there are $3M-1$ matrices.

The extension to qudits (i.e. local $d\times d$ matrices) is equally straightforward \cite{Br}.

\paragraph{8. Universal gates from the $R$-matrix.}

Straightforward option for any kind of a quantum computer would be to just
realize the universal gates, and then the rest is left to the algorithm writers,
who are used to reduce every algorithm to a sequence of the universal gate operations
(and their multi-qubit analogues).

Let us first note that one can immediately realize the single qubit operations (\ref{Hadamard}) and (\ref{Pgate}) by the matrices ${\cal R}$ and ${\cal S}$ or  $\bar {\cal R}$ and $\bar {\cal S}$ from (\ref{R}), (\ref{S}). For the first pair, one can just put $q=e^{-3\pi i/8}$ and $A=e^{-3\pi i/2}$, which corresponds to the group $SU(4)$. In fact, one evidently can consider any higher $N$ multiple 4: $N=4k$ and choose $q=e^{-3\pi i/(8k)}$.
For the second pair, it is described, e.g. by the pair $q=e^{\pi i/3}$ and $A=e^{5\pi i/4}$, which is less plausible corresponding to the non-integer $N=15/4$, however, is technically possible. Note that for fault-tolerant constructions one also usually includes in the universal set ${\cal R}^2$ as well \cite{Nielsen}. The two elements ${\cal R}$ and ${\cal R}^2$, indeed, look completely different from the point of view of knots.

As we explained in paragraph 6, one needs in practice the combinations $HTHT$ and $THTH$. Taking into account associating ${\cal R}$ and ${\cal S}$ with $T$ and $H$ respectively, we immediately notice that

\be
HTHT={\cal R}_2{\cal R}_1,\ \ \ \ \ \ \ \ \ \ THTH={\cal R}_1{\cal R}_2
\ee
since ${\cal R}_1$ is related with ${\cal R}_2$ exactly by the rotation with the Racah matrix ${\cal S}$. Thus, we finally come to the claim that the two $R$-matrices ${\cal R}_1$ and ${\cal R}_2$ describing the two-bridge block in the plat representation (in this representation of knots, ${\cal R}_1={\cal R}_3$) provide the local gates dense in the space of single qubit unitary matrices. Similarly, one can work with the matrices $\bar {\cal R}$ and $\bar {\cal S}$ from (\ref{R}), (\ref{S}). In fact, in order to generate all single qubit gates, one can choose almost arbitrary unimodular $q$ and $A$ preserving matrices' unitarity, the values discussed above are just demonstrate that one reproduces the standard universal set of gates.

Now, as it was explained in the previous paragraph, in order to construct a set of universal gates, it is necessary and sufficient to add any entangling matrix. As a matter of fact, typical (i.e. at $q\ne 1$) $R$-matrices and Racah matrices are entangling (see examples in \cite{Kauff3}). Hence, to generate the set of universal gates in the two-qubit case, it is sufficient to consider intertwined pairs of two-bridge knots/links (see Fig.4): they are intertwined by the needed entangling matrices. In fact, the construction of arbitrary unitary matrices in the previous paragraph used as convenient entangling matrices the CNOT matrices $K_i$ which act on two neighbour factors in the tensor product. It is much similar to the entangling matrices which act between two neighbour two-bridge braids, which present, in this way, a realization of separate qubits.

\begin{figure}[h]
  \centering
\includegraphics[width=10cm]{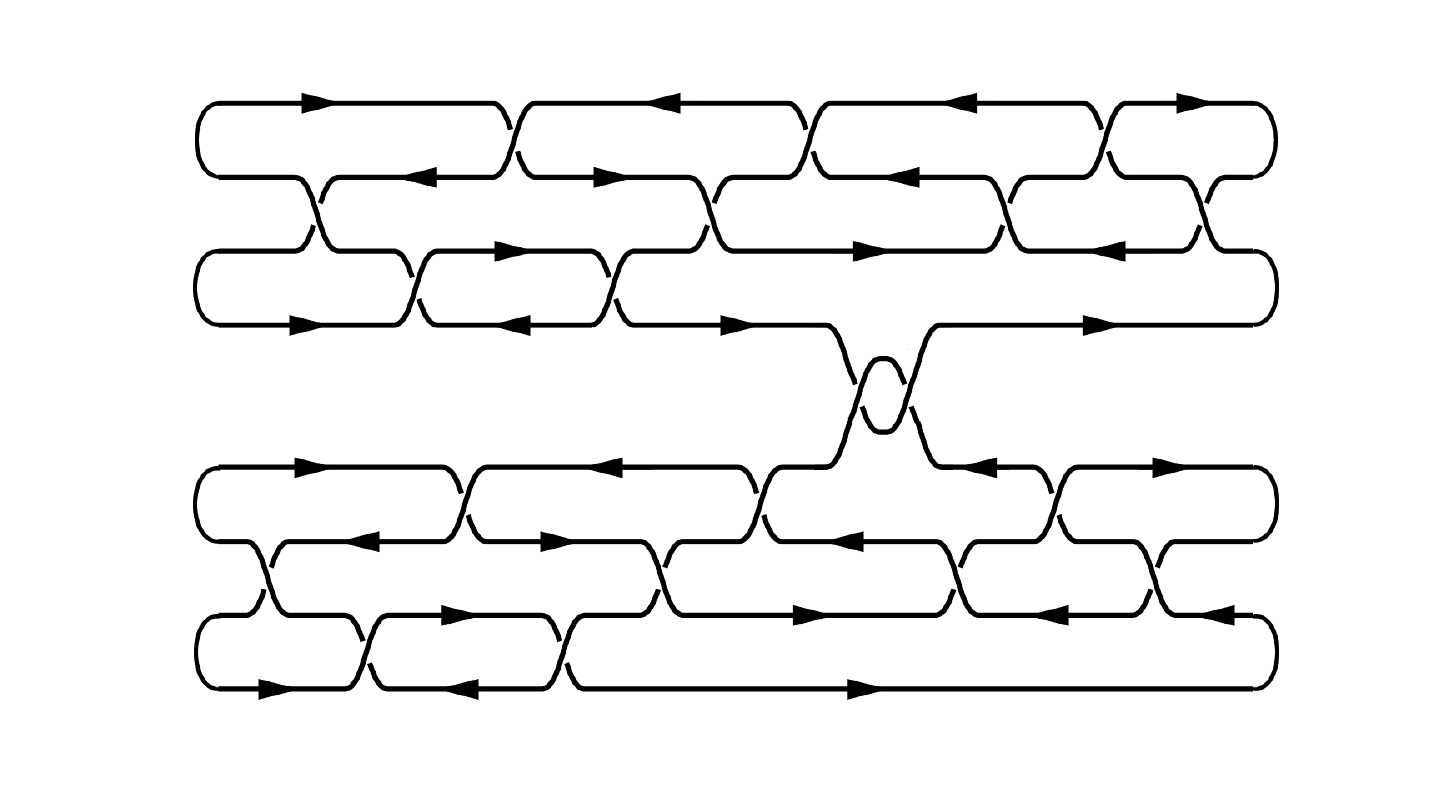}
\caption{The plat representation corresponding to two-qubits: four-bridge case.}
  \label{fig:4}
\end{figure}

Thus, in this picture, {\bf one qubit is realized by a two-bridge block, while for higher number of qubits, the number of two-bridge blocks in the knot/link has to be correspondingly enlarged. Further, one could generate qudits considering higher (non-fundamental) group representations} (as we explained above, two matrices are still sufficient in this case to generate all $d\times d$ unitary matrices). For instance, considering the $r$-th symmetric representations, what is convenient, since they exist for any group, one can expect generating qudits with $d=r+1$.

In fact, the standard universal set of gates considered above is in no way distinguished: there are many other examples of universal gates, with their own advantages and drawbacks. More important is the fact that using these universal gates one may need exponentially many gates to approximate some unitary operations (see, e.g., \cite[s.4.5.4]{Nielsen}).
Hence, one may ask for a wider and more effective set of elementary gates. One may address a fundamental question:
{\bf can universal gates be generated by the ${\cal R}$-matrices}?
As we explained above, {\it their} products are at least expected to be dense in the space of all unitary matrices.

We emphasize that the ${\cal R}$-matrices provide a {\it new} systematic approach to quantum programming, which is not just a literal reformulation of the previously developed algorithms. The differences become more pronounced when one proceeds from qubits to qudits:  for ${\cal R}$-matrices this just involves higher representations, while the knots (codes) can remain the same, while in naive approaches one often needs to use absolutely new collections of universal gates for different spins. A counterpart of this ambiguity is existence of Reidemeister equivalent combinations of ${\cal R}$-matrices, which implies that one and the same knot/code can be described by different combinations of ${\cal R}$, $\bar{\cal R}$, ${\cal S}$ and $\bar {\cal S}$.

\paragraph{9. Entanglement of operators: quantum vs. topological entanglement.}

As we already explained in paragraph 6, in accordance with the Brylinskis theorem the universal gate should contain, at least, one entangling operator \cite{Br}. It turns out that there exists much similar theorem in knot theory, which allows one to put the similarity between quantum and topological entanglement (that was proposed long ago \cite{Ara,Kauf}) on a quantitative ground. The theorem \cite{ten} claims that the knot polynomial can distinguish different knots only if it is constructed with the help of $R$-matrix which is an entangling operator. In particular, this means that, using the full permutation matrix (which is not entangling) one can produce only trivial knot polynomials. This is really the case, since it would produce the HOMFLY polynomial at $q=1$, which is just 1.

In variance with full permutation matrix, the $R$-matrix at $q\ne 1$ is an entangling operator. However, there are different degrees of entanglement, and one can introduce quantitative description of the entangling power of the operator. The simplest one is a degree of entanglement that measures enlarging the rank of the vector with the action of operator on it: what is the minimal number of terms in  $\hat X|v>$, where $|v>=|x>\otimes |y>$ runs over the whole variety of vectors $|x>$ and $|y>$. However, besides the discrete degree, there should be continuous characteristics (like Vandermonde determinant  for the matrix of the maximal rank). In principle, they should be looked among the objects of (non-)linear algebra \cite{NLA} and matrix model theory, which can often be put in the entropy like form.

In order to define such more subtle characteristics of entanglement power of operators, one has to consider first the entanglement of states:
consider a state of a system consisting of two parts $A$ and $B$, it can be
represented as
$
\Psi = \sum_{a,i} c_{ai} \,|a\rangle\otimes |i\rangle
$\ 
with rectangular matrix $c_{ai}$ and normalization condition (where different signs of trace denote the traces in $a$ and $i$)\
$
\Tr (cc^\dagger) = \tr (c^\dagger c) = 1
$. 
Its AB-entanglement can be measure by the entanglement entropy
$
S_{AB}(\Psi) = \Tr \Big(c^\dagger c \log c^\dagger c\Big) =\tr \Big(c^\dagger c\log c^\dagger c\Big)
$, 
i.e. $S_{AB}(\Psi) = S_{BA}(\Psi)$.
The entanglement entropy vanishes for {\it pure} ($AB$-pure) states,
when rank of the matrix $c$ is one.

Entanglement concerns operators which act on linear spaces with an additional structure, namely, which are tensor products of constituent spaces. Such operators can be conveniently decomposed as
$
\hat X = \sum_l s_l \hat A_l\otimes \hat B_l
$\ 
in the orthonormal basis of $\hat A$ and $\hat B$. For such Schmidt-decomposed operators, one can define various quantities that describe the entangling power of operators:
\begin{itemize}
\item[a)]
The most immediate one
$
{\cal S}(X) = \sum_l s_l^2\log s_l^2
$.
\item[b)]
One can also look at the number of negative eigenvalues
$
{\cal N}(X) = \# \{s_l<0\}
$.
\item[c)]
There are also minimization characteristics,
over all pure states:
$
{\rm min}_{\psi,\chi} S_{AB}\Big(\hat X(|\psi>\otimes|\chi>)\Big)
$\ 
or over all states:
$
{\rm min}_{\Psi} \left(S_{AB}\Big(\hat X(\Psi)\Big)-S_{AB}\Big(\Psi\Big)\right)
$.
\item[d)] At last, there are averaged characteristics like the entanglement power of \cite{perms}:
$
\epsilon(\hat X) = \int \frac{d}{d-1}\Big(1-\Tr \rho(\hat X)^2\Big)\ d\psi d\chi
$\ 
for $\rho(\hat X) =
\hat X\Big(|\psi>\otimes|\chi> <\psi|\otimes <\chi|\Big)\hat X^\dagger$.
\end{itemize}

This list is neither distinguished, nor complete. A search of adequate definitions and criteria is still in progress, and the group theory structures introduced through the use of quantum $R$-matrices should be very helpful in this respect as well.

\paragraph{10. Physical realizations.}

This is a separate big issue of principal importance. It deserves noting that the very idea of {\it topological} computing appeared from the very beginning as a possible tool for diminishing the loss of quantum coherence, and in this quality it remains in the center of interest in practical attempts to build a working quantum computer device \cite{Kitaev}. The problem, however, is that the physical systems with non-Abelian anyonic statistics are still exotic in solid state physics. Thus, it is somewhat premature to go into details about the next step: how the entire variety of quantum ${\cal R}$-matrices, for different groups and representations can be physically realized, especially, if one wants to leave all the three variables, $q$, $A=q^N$ and representation $R$  as independent free parameters. This is a great and important challenge for different branches of solid state physics (today it is mostly studied in relation with the quantum Hall effect, but mostly because it is also a story with a "topological" flavor). Here we add just a few simple remarks.

As we saw in the previous paragraphs, the unitary evolution was provided by the $R$-matrices acting in the space of intertwining operators. In other words, this means that one has to deal with non-standard spin systems (or spin chains) with the energy that depends only on the full moments of pairs of neighbour spins. For instance, if one considers a pair of spins in the fundamental representations $[1]$ of the $SU(N)$ spin chain, the energies of two possible representations can be parameterized as $E_{[2]}=\epsilon$ and $E_{[11]}=\pi-\epsilon$ so that the evolution matrix is
\be
\hat U=\left(\begin{array}{cc}
q&0\\
0&-{1\over q}
\end{array}\right)
\ee
with $q=e^{i\epsilon t}$ (in the concrete Chern-Simons realization one still can consider, say, $t=1$, which gives rise to a discrete evolution, see \cite{evo}, and $\epsilon=2\pi/(\kappa+N)$) .

This evolution matrix is diagonal, since we constructed it in the basis of irreps in the product of two spins. Having further three spins, in the case of the evolution operation acting on the second pair of spins, one should recalculate to the basis, and this is done exactly by the Racah matrices. Now, one has to consider higher representations and more spins, which, as we described in this note, is related to knot polynomials. This gives a physical realization of the construction.

Another possibility is still to use the $R$-matrix acting in the product of representations, like (\ref{Rf}). This $R$-matrix is not unitary, but it is Hermitean at real $q$. This means that exponential of this $R$-matrix is a unitary operator, and one can consider the corresponding evolution. This leads not to the standard spin chain, since its Hamiltonian is equal to
\be
H=\sum_i R_{i,i+1}
\ee
where the sum runs over spins, so that the evolution operator is
\be
\hat U=e^{iHt}
\ee
instead of
\be
\prod_i e^{iR_{i,i+1}t}
\ee
However, this system is in the same class of well-studied spin chains. Still, this system is not that immediately related to knot polynomials, in variance with the previous variant.

\paragraph{11. Conclusion.}

In this letter we developed the earlier ideas
and suggested to treat {\it topological} computing as a distinguished
and standard setting version of quantum programming.
For this to work, one should be able to produce universal gates from the
${\cal R}$-matrices acting in the space of intertwining operators.
{\bf The hope is that arbitrary exponentially large $d^M\times d^M$ unitary matrices
can be approximated by products of linearly many $\sim M$ ${\cal R}$-matrices}, acting on adjacent
two-bridge braids in the multi-bridge (plat) representation of knots/links
(parameter $d$ is, roughly, size of the representation $R$).
{\bf If true, this should be a topological version of the Brylinskis theorem.}
Note that, like in that theorem, the number of these matrices (universal gates)
is exponentially smaller than the number of Chevalley generators of $U_{d^M}$.
As to irrationality needed for such an approximation, it comes from the
eigenvalues of the diagonal matrices like ${\cal R}$ and $\bar {\cal R}$ and in ${\cal R}$-matrix
theory is regulated by the second Casimir operator eigenvalues of relevant representations.
{\bf If true, this construction would identify amplitudes generated by the topological
computer with the knot polynomials, which are {\it polynomials} with {\it integer}
coefficients, and this additional integrality can be used for fault-tolerant constructions.}
For knot theory, a look from this perspective enlightens new amusing properties
of knot polynomials like the restriction $|H_R/d_R|\leq 1$ for unimodular $q$
(and restricted $N$).
In extreme, the $NP$-problem of knot polynomial calculation can appear partly solvable
not only by quantum computers, but also pure theoretically (by exploiting
integrability and other deep properties of the theory).
{\bf If true, topological computing implies classification of quantum algorithms
by knots}, and it becomes then a challenging problem to decide, which knot polynomial
needs to be calculated to solve, say, a prime decomposition problem,
and whether it is hopeless in modern knot theory or not.
The parallel search for physical realization of topological computer,
which is quite non-trivial, especially for arbitrary (or at least unimodular)
values of parameters $q$ and $A$,
and for an exhaustive theory of knot polynomials should benefit both sides.

\paragraph{Acknowledgements.}

The authors thank the International Institute of Physics in Natal for hospitality.
Our work was partly supported by the Brazilian Ministry of Education and by grant
16-32-60047-Mol-a-dk (And.Mor), by RFBR grants 16-01-00291 (D.Mel., A.Mir.), 15-01-05990 (S.Mir.), 16-02-01021 (A.Mor.),
17-01-00585 (An.Mor.), by joint grants
17-51-50051-YaF, 15-51-52031-NSC-a, 16-51-53034-GFEN,
16-51-45029-IND-a.


\begin{thebibliography}{12}

\bibitem{Nielsen} M.A. Nielsen and I.L. Chuang, {\sl Quantum Computation and Quantum Information}, Cambridge University Press, 2000

\bibitem{QC} A.Yu. Kitaev, A.H. Shen and M.N. Vyalyi, {\sl Classical and quantum computation}, Providence, RI: AMS, American Mathematical Society. xiii, 257 pp. (2002) [Graduate Studies in Mathematics, 47]\\
M. Hayashi, S. Ishizaka, A. Kawachi, G. Kimura and T. Ogawa, {\sl Introduction to Quantum Information Science}, Springer, 2015

\bibitem{Nayak} Ch. Nayak, S.H. Simon, A. Stern, M. Freedman and S.D. Sarma, Rev. Mod. Phys. {\bf 80} (2008) 1083,	arXiv:0707.1889

\bibitem{NL} J. von Neumann, {\sl Theory of Self-Reproducing Automata}, Univ. of Illinois Press, 1966

\bibitem{Kauf} L.Kauffman, S.Lomonaco, New Journal of Physics, {\bf 4} (2002) 73.1-18; {\bf 6} (2004) 134.1-40, quant-ph/0401090

\bibitem{RTmod} A. Mironov, A. Morozov and An. Morozov, JHEP {\bf 03} (2012)
034, arXiv:1112.2654\\
H. Itoyama, A. Mironov, A. Morozov, An. Morozov,
Int.J.Mod.Phys. {\bf A27} (2012) 1250099,
arXiv:1204.4785\\
A. Anokhina, A. Mironov, A. Morozov and An. Morozov, Nucl.Phys. {\bf B868} (2013) 271-313,
arXiv:1207.0279

\bibitem{inds} R.K. Kaul and T.R. Govindarajan, Nucl.Phys. {\bf B380} (1992)
293-336, hep-th/9111063\\
P. Ramadevi, T.R. Govindarajan and R.K. Kaul, Nucl.Phys. {\bf B402} (1993)
548-566, hep-th/9212110;
Nucl.Phys. {\bf B422} (1994) 291-306, hep-th/9312215\\
P. Ramadevi and T. Sarkar,
Nucl.Phys. B600 (2001) 487-511,
hep-th/0009188\\
Zodinmawia and P. Ramadevi, arXiv:1107.3918;  arXiv:1209.1346

\bibitem{Gar} R.Gelca, Math. Proc. Cambridge Philos. Soc. {\bf 133} (2002)
311-323,
math/0004158;\\
R.Gelca and J.Sain, J. Knot Theory Ramifications, {\bf 12} (2003) 187-201,
math/0201100;\\
S.Gukov, Commun.Math.Phys. {\bf 255} (2005) 577-627, hep-th/0306165;\\
S.Garoufalidis, Geom. Topol. Monogr. 7 (2004) 291-309,  math/0306230

\bibitem{IMMMfe} H. Itoyama, A. Mironov, A. Morozov and An. Morozov,
JHEP {\bf 2012} (2012) 131,  arXiv:1203.5978;  IJMP {\bf A27} (2012) 1250099,  arXiv:1204.4785

\bibitem{MMpol} A.Mironov and A.Morozov,  AIP Conf.Proc. {\bf 1483} (2012) 189-211, arXiv:1208.2282

\bibitem{MMMI} A. Mironov, A. Morozov and An. Morozov, {\sl Strings, Gauge Fields, and the Geometry Behind:
The Legacy of Maximilian Kreuzer,} World Scietific Publishins Co.Pte.Ltd. 2013, pp.101-118, arXiv:1112.5754

\bibitem{AG} L. Alvarez-Gaune, C. Gomez and S. Sierra, Phys.Lett. {\bf B220} (1989) 142

\bibitem{Freedman} M. Freedman, 
Proc. Natl. Acad. Sci., USA, {\bf 95} (1998), 98–101\\
M.Freedman, A.Kitaev, M.Larsen, Z.Wang, 
Bull.Amer.Math.Soc. (N.S.) {\bf 40} (2003) 31–38, quant-ph/0101025\\
M. H. Freedman, M. Larsen and Z. Wang, 
Commun.Math.Phys. {\bf 227} (2002) 605-622, quant-ph/0001108

\bibitem{knotpols} J.W. Alexander, 
Trans.Amer.Math.Soc. {\bf 30} (2) (1928) 275-306\\
V.F.R. Jones, 
Invent.Math. {\bf 72} (1983) 1
Bull.AMS {\bf 12} (1985) 103
Ann.Math. {\bf 126} (1987) 335\\
L. Kauffman,
Topology {\bf 26} (1987) 395\\
P. Freyd, D. Yetter, J. Hoste, W.B.R. Lickorish, K. Millet,
A. Ocneanu,
Bull. AMS. {\bf 12} (1985) 239\\
J.H. Przytycki and K.P. Traczyk, 
Kobe J. Math. {\bf 4} (1987) 115-139
J.H. Conway, 
Algebraic Properties,
In: John Leech (ed.), {\sl Computational Problems in Abstract Algebra}, Proc.
Conf.
Oxford, 1967, Pergamon Press, Oxford-New York, 329-358, 1970

\bibitem{Wit} E. Witten,
Comm.Math.Phys. {\bf 121} (1989)  351-399

\bibitem{RT} E. Guadagnini, M. Martellini, M. Mintchev, Clausthal 1989,
Procs.
 307-317;
Phys.Lett. {\bf B235} (1990) 275\\
N.Yu. Reshetikhin and V.G. Turaev, 
Comm. Math. Phys. {\bf 127} (1990) 1-26

\bibitem{Ara} P.K. Aravind, 
in: {\sl Potentiality, Entanglement and Passion-at-a-Distance}, ed. by R.S. Cohen et al, pp. 53-59, Kluwer, 1997

\bibitem{Hecke} D. Goldschmidt, {\sl Group Characters, Symmetric Functions, and the Hecke Algebra}, University Lecture Series, {\bf 4}, American Mathematical Society, Providence, RI, 1993

\bibitem{TL} L.H. Kauffman,
Topology, {\bf 26(3)} (1987) 395-407

\bibitem{Kauff2} L.H. Kauffan, math/0105255; arXiv:1301.6214\\
D.~Aharonov, V.F.R.~Jones and Z.~Landau,
quant-ph/0511096

\bibitem{ten} G. Alagic, M. Jarret and S.P. Jordan, J. Phys. A: Math. Theor. {\bf 49} (2016) 075203, arXiv:1507.05979\\
L.H. Kauffman and E. Mehrotra, arXiv:1611:08047

\bibitem{YB} V. Chari and A. Pressley, {\sl A Guide to Quantum Groups}, (1994), Cambridge University Press, Cambridge\\
J. Fuchs, {\sl Affine Lie Algebras and Quantum Groups}, (1995), Cambridge University Press, Cambridge

\bibitem{sch} R.J. Baxter, {\sl Exactly solved models in statistical mechanics}, London, Academic Press, 1982

\bibitem{KBI}
V.E. Korepin, N.M. Bogoliubov and A.G. Izergin, {\sl Quantum Inverse Scattering Method and Correlation Functions}, (1997), Cambridge University Press, Cambridge

\bibitem{loop} A.V. Razumov and Yu.G. Stroganov, J.Phys. {\bf A34} (2001) 3185,
cond-mat/0012141\\
M.T. Batchelor, J. de Gier and B. Nienhuis, J.Phys. {\bf A34} (2001) L265-L270, cond-mat/0101385

\bibitem{KhR} M. Khovanov, Duke Math.J. {\bf 101} (2000) no.3, 359426, math/9908171;
Experimental Math. {\bf 12} (2003) no.3, 365374, math/0201306;
J.Knot theory and its Ramifications {\bf 14} (2005) no.1, 111-130, math/0302060;
Algebr. Geom. Topol. {\bf 4} (2004) 1045-1081, math/0304375;
Int.J.Math. {\bf 18} (2007) no.8, 869885, math/0510265; math/0605339; arXiv:1008.5084\\
D. Bar-Natan, Algebraic and Geometric Topology {\bf 2} (2002) 337-370, math/0201043;
 Geom.Topol. {\bf 9} (2005) 1443-1499, math/0410495;
 J.Knot Theory Ramifications {\bf 16} (2007) no.3, 243255, math/0606318

\bibitem{DM} V. Dolotin and A. Morozov, JHEP {\bf 1301} (2013) 065, arXiv:1208.4994; J. Phys. {\bf 411} 012013,
arXiv:1209.5109; Nucl.Phys. {\bf B878} (2014) 12-81,  arXiv:1308.5759

\bibitem{J} M. Jimbo, Lett. Math. Phys. {\bf 10} (1985) 63-69

\bibitem{DIM} H. Awata, H. Kanno, A. Mironov, A. Morozov, An. Morozov, Y. Ohkubo, Y. Zenkevich, arXiv:1611.07304

\bibitem{CS} S.-S. Chern and J. Simons,
Ann.Math. {\bf 99} (1974) 48-69

\bibitem{CFTKM} E. Witten, Comm.Math.Phys. {\bf 92} (1984) 455\\
A.M. Polyakov and P.B. Wiegmann, Phys.Lett. {\bf B131} (1983) 121\\
V. Knizhnik and A. Zamolodchikov, Nucl.Phys. {\bf B247} (1984) 83-103\\
D. Gepner and E. Witten, Nucl.Phys. {\bf B278} (1986) 493

\bibitem{CSmore} G.Moore and N.Seiberg, Phys.Lett. B220 (1989) 422;\\
V.Fock and Ya.I.Kogan, Mod.Phys.Lett. A5 (1990) 1365-1372;\\
R.Gopakumar and C.Vafa, Adv.Theor.Math.Phys. 3 (1999) 1415-1443, hep-th/9811131

\bibitem{AS} M.Aganagic and Sh.Shakirov, arXiv:1105.5117

\bibitem{GMM} D.Galakhov, A.Mironov, A.Morozov,
JETP, {\bf 120} (2015) 549-577 (ZhETF, {\bf 147} (2015) 623),  arXiv:1410.8482\\
D. Galakhov, D. Melnikov, A. Mironov and A. Morozov, Nucl.Phys. {\bf B899} (2015) 194-228, arXiv:1502.02621

\bibitem{katlas} D.Bar-Natan, http://www.katlas.org

\bibitem{Br} J.L. Brylinski and R. Brylinski, in: {\sl Mathematics of Quantum Computation}, eds. R. Brylinski and G. Chen, (Chapman \& Hall/CRC Press, Boca Raton, Florida, 2002), quant-ph/0108062

\bibitem{Kauff3} Y. Zhang, L.H. Kauffman and M.-L. Ge, International Journal of Quantum Information, {\bf 3} (2005) 669-678, quant-ph/0412095; Quant. Inf. Proc. {\bf 4} (2005) 159-197, quant-ph/0502015

\bibitem{NLA} V.Dolotin and A.Morozov, {\sl Introduction to Non-Linear Algebra},
World Scientific, 2007, hep-th/0609022

\bibitem{perms} L. Clarisse, S. Ghosh, S. Severini and A. Sudbery, Phys. Rev. {\bf A72} (2005) 012314, quant-ph/0502040

\bibitem{evo} A. Mironov, A. Morozov and An. Morozov, AIP Conf. Proc. {\bf 1562} (2013) 123, arXiv:1306.3197;
 Mod. Phys. Lett. {\bf A 29} (2014) 1450183,  arXiv:1408.3076\\
 S. Arthamonov, A. Mironov, A. Morozov and An. Morozov,  JHEP {\bf 04} (2014) 156,  arXiv:1309.7984\\
 A. Mironov, A. Morozov and A. Sleptsov, JHEP {\bf 07} (2015) 069,  arXiv:1412.8432

\bibitem{Kitaev} A.Y. Kitaev,
Ann. of Phys. {\bf 303} (2003) 2–30, quant-ph/9707021\\
M.H. Freedman, A. Kitaev, Z. Wang, 
Commun.Math.Phys. {\bf 227} (2002) 587-603, quant-ph/0001071

\end{thebibliography}
\end{document}